\documentclass[aps,prl,twocolumn,showpacs,preprintnumbers,amsmath,amssymb, superscriptaddress, bibnotes,longbibliography]{revtex4}
\usepackage{graphicx}
\usepackage{dcolumn}
\usepackage{bm}
\usepackage[colorlinks,linkcolor=blue,anchorcolor=blue,citecolor=blue,urlcolor=blue,filecolor=blue,menucolor=blue,runcolor=blue]{hyperref}
\usepackage{ulem}

\makeatletter
\@ifundefined{textcolor}{}
{%
	\definecolor{BLACK}{gray}{0}
	\definecolor{WHITE}{gray}{1}
	\definecolor{RED}{rgb}{1,0,0}
	\definecolor{GREEN}{rgb}{0,1,0}
	\definecolor{BLUE}{rgb}{0,0,1}
	\definecolor{CYAN}{cmyk}{1,0,0,0}
	\definecolor{MAGENTA}{cmyk}{0,1,0,0}
	\definecolor{YELLOW}{cmyk}{0,0,1,0}
}

\usepackage{color}

\newcommand{\beq}{\begin{eqnarray}}
\newcommand{\eeq}{\end{eqnarray}}
\newcommand{\ys}[1]{\textcolor{black}{#1}}
\newcommand{\Neel}{N\'eel}

\makeatother
\usepackage[english]{babel}

\begin{document}

\title{Quasi-one-dimensional spin excitations in the iron pnictide NaFe$_{0.53}$Cu$_{0.47}$As}

\author{Yifan Wang}
\affiliation{Center for Correlated Matter and School of Physics, Zhejiang University, Hangzhou 310058, China}

\author{David W. Tam}
\affiliation{Department of Physics and Astronomy, Rice University, Houston, Texas 77005, USA}

\author{Weiyi Wang}
\affiliation{Department of Physics and Astronomy, Rice University, Houston, Texas 77005, USA}

\author{R. A. Ewings}
\affiliation{ISIS Pulsed Neutron and Muon Source, STFC Rutherford Appleton Laboratory, Harwell Campus, Didcot, Oxon, OX11 0QX, United Kingdom}

\author{J. Ross Stewart}
\affiliation{ISIS Pulsed Neutron and Muon Source, STFC Rutherford Appleton Laboratory, Harwell Campus, Didcot, Oxon, OX11 0QX, United Kingdom}

\author{Masaaki Matsuda}
\affiliation{Neutron Scattering Division, Oak Ridge National Laboratory, Oak Ridge, Tennessee 37831, USA}

\author{Chongde Cao}
\affiliation{Department of Physics and Astronomy, Rice University, Houston, Texas 77005, USA}

\author{Changle Liu}
\affiliation{School of Physics and Beijing Key Laboratory of Opto-electronic Functional Materials and Micro-nano Devices, Renmin University of China, Beijing 100872, China}

\author{Rong Yu}
\email{rong.yu@ruc.edu.cn}
\affiliation{School of Physics and Beijing Key Laboratory of Opto-electronic Functional Materials and Micro-nano Devices, Renmin University of China, Beijing 100872, China}
\affiliation{Key Laboratory of Quantum State Construction and Manipulation (Ministry of Education), Renmin University of China, Beijing, 100872, China}

\author{Pengcheng Dai}
\email{pdai@rice.edu}
\affiliation{Department of Physics and Astronomy, Rice University, Houston, Texas 77005, USA}
\affiliation{Rice Laboratory for Emergent Magnetic Materials and Smalley-Curl Institute, Rice University, Houston, TX 77005, USA}

\author{Yu Song}
\email{yusong\_phys@zju.edu.cn}
\affiliation{Center for Correlated Matter and School of Physics, Zhejiang University, Hangzhou 310058, China}

\begin{abstract}
Spectroscopic measurements in model one-dimensional (1D) correlated systems offer insights for understanding their two-dimensional counterparts, which include the cuprate and iron pnictide/chalcogenide superconductors. A major challenge is the identification of such correlated systems with dominantly 1D physics. In this work, inelastic neutron scattering measurements on NaFe$_{0.53}$Cu$_{0.47}$As single crystal directly reveal quasi-1D spin excitations, resulting from atomic order that lead to magnetic Fe and nonmagnetic Cu chains. The dominant exchange interaction is antiferromagnetic along the chain ($SJ_{\rm \parallel}\approx90.1(3)$~meV), whereas the inter-chain couplings are much weaker ($SJ_{\rm \perp}\approx-2.4(1)$~meV and $SJ_{\rm c}\approx0.15(5)$~meV). The quasi-1D spin excitations in NaFe$_{0.53}$Cu$_{0.47}$As stem from both the {\Neel} and stripe vectors, with {\Neel} excitations sensitive to Fe impurities on the Cu site. The spin excitations in quasi-1D NaFe$_{0.53}$Cu$_{0.47}$As and quasi-2D FeSe exhibit a striking resemblance, suggesting a common origin for their coexistent stripe and {\Neel} excitations. Our findings demonstrate magnetic dilution in NaFeAs leads to dimension reduction of its magnetic degree of freedom, presenting a strategy for discovering low-dimensional quantum materials.    
\end{abstract}

\maketitle

Understanding the behaviors of correlated electrons in the cuprates and the iron pnictides/chalcogenides (FePn/Ch) is a central problem in condensed matter physics \cite{Lee2006,Keimer2015,Si2016,Fernandes2022}. The discovery of superconducting ladder cuprates \cite{Uehara1996,Nagata1998} and FePn/Ch \cite{Takahashi2015,Yamauchi2015,JYing2017} indicate that the essential physics of these systems are retained approaching the one-dimensional (1D) limit, and suggest quasi-1D materials as model systems for studying correlated electrons and their emergent properties.  
In the cuprates, spectroscopic studies of 1D systems such as chains and ladders can be benchmarked against numerical solutions of many-body Hamiltonians \cite{Dagotto1996,Walters2009,chain_Cuprate_Science}, and further led to observations such as spin-charge separation \cite{Schlappa2012,Kim2006}, multi-spinons \cite{Schlappa2018}, and long-range quantum entanglement in the solid state \cite{Sahling2015}. 

Compared to the cuprates, the multi-orbital nature of the FePn/Ch gives rise to orbital-selectivity absent in the cuprates \cite{Yi2013,Yu2013,YLi2016,Sprau2017,YSong2018,Yu2021}, leading to rich phase diagrams and magnetic phases in 1D systems \cite{Bronger1987,Nishi1992,QLuo2014,Pandey2020,Zhang2021,Caron2011,Caron2012,Nambu2012,Du2012,Herbrych2018,Herbrych2019,Herbrych2021,Wu2019}. For instance, since a two-orbital chain maps to a single-orbital ladder, a multi-orbital chain could behave like a single-orbital ladder system and realize a superconducting ground state \cite{Patel2017,Jiang2018,Patel2020,TlFeSe2_SC2020}. However, due to a scarcity of 1D FePn/Ch, there have been limited experimental studies of such physics. Moreover, the crystal morphology of typical 1D FePn/Ch hinders spectroscopic measurements of their electronic structures via scanning tunneling microscopy (STM) or angle-resolved photoemission (ARPES), and measurements of their spin excitations have mostly been limited to powder samples \cite{Mourigal2015,MWang2016,MWang2017,RbFeS2,Zliu2022}. 


The NaFe$_{1-x}$Cu$_x$As (NFCA) series provides another route for \ys{realizing FePn/Ch in} the 1D limit. For $x\sim2\%$, the spin-density-wave in NaFeAs is suppressed giving way to a superconducting dome with maximal $T_{\rm c}\approx10$~K , similar to most FePn superconductors \cite{AFWang2013,GTan2017}. With increasing Cu-substitution the system becomes semiconducting/insulating for $x\gtrsim0.1$, accompanied by the development of short-range Fe-Cu atomic order and antiferromagnetic (AF) order, both of which approach long-range when $x\approx0.5$ \cite{AFWang2013,CYe2015,Song2016,YXin2020}. The ideal NFCA with $x=0.5$ (NFCA50) is a correlated insulator \cite{Song2016,CEMatt2016,ACharnukha2017,SZhang2017}, with Fe-Cu ordering leading to AF Fe chains separated by nonmagnetic Cu chains \cite{Song2016}. While these atomic chains should lead to quasi-1D electronic states and magnetism, ARPES measurements find a two-dimensional (2D) electronic structure near the Fermi level \cite{CEMatt2016}, and the overlapping of scattering from different domains at small momenta hampered direct spectroscopic observations of quasi-1D magnetism via resonant inelastic X-ray scattering (RIXS) \cite{YSong2021}. 

\begin{figure}
	\includegraphics[angle=0,width=1\columnwidth]{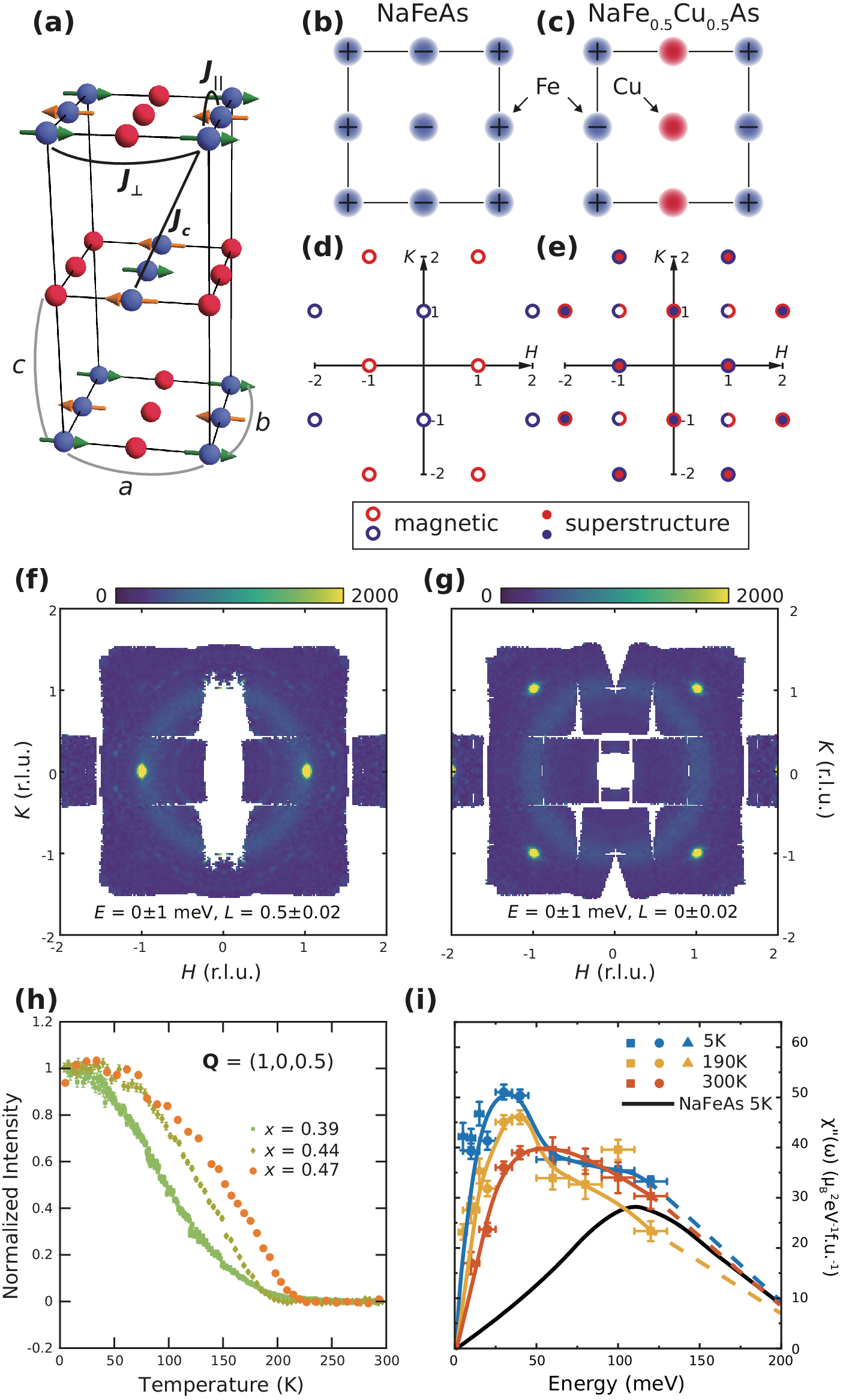}
	\vspace{-12pt} \caption{(a) Schematic crystal and magnetic structures of NFCA50. In-plane magnetic structures for (b) NaFeAs and (c) NFCA50. These structures \ys{(blue symbols)} and their 90$^{\circ}$-rotated twins \ys{(red symbols)} lead to magnetic and superstructure peaks in (d) and (e), where the main Bragg peaks are not shown. Elastic scattering data for NFCA47 in the (f) $L=0.5$ and the (g) $L=0$ planes. The data are folded into a single quadrant and reproduced in other quadrants. (h) Temperature dependence of the normalized elastic magnetic scattering in NFCA. The $x=0.47$ data are measured with polarized neutrons \cite{SI}. The $x=0.39$ and 0.44 data are from Ref.~\cite{Song2016}, measured with unpolarized neutrons. (i) The local susceptibility $\chi(\omega)''$ of NFCA47, compared with that of NaFeAs at 5~K \cite{CZhang2014}. The square, circle and triangle symbols are for $E_{\rm i}=50$~meV, 81~meV, and 245~meV, respectively. The solid lines are guides-to-the-eye.}
	\vspace{-12pt}
	\label{Schematic}
\end{figure}

In this work, we report spin excitations in the FePn chain NFCA with $x=0.47$ (NFCA47, approximating NFCA50) measured via inelastic neutron scattering. The spin excitations reveal a prominent anisotropy of exchange interactions for directions along and perpendicular to the Fe chains, directly demonstrating the quasi-1D character of magnetism. Strikingly, the magnetic scattering patterns of NFCA47 exhibit remarkable similarities to those in FeSe \cite{Wang2016,RLiu2024}. Common to the two systems are spin excitations that stem from both stripe and {\Neel} vectors of the square lattice, a strong in-plane anisotropy, and \ys{an effective moment} $S\approx1$. These findings demonstrate NFCA50 to be a quasi-1D insulating FePn with magnetic excitations resembling superconducting FeSe, and \ys{showcase magnetic dilution as a viable approach for discovering quasi-1D systems in layered compounds.}  

\begin{figure*}
	\includegraphics[angle=0,width=1\textwidth]{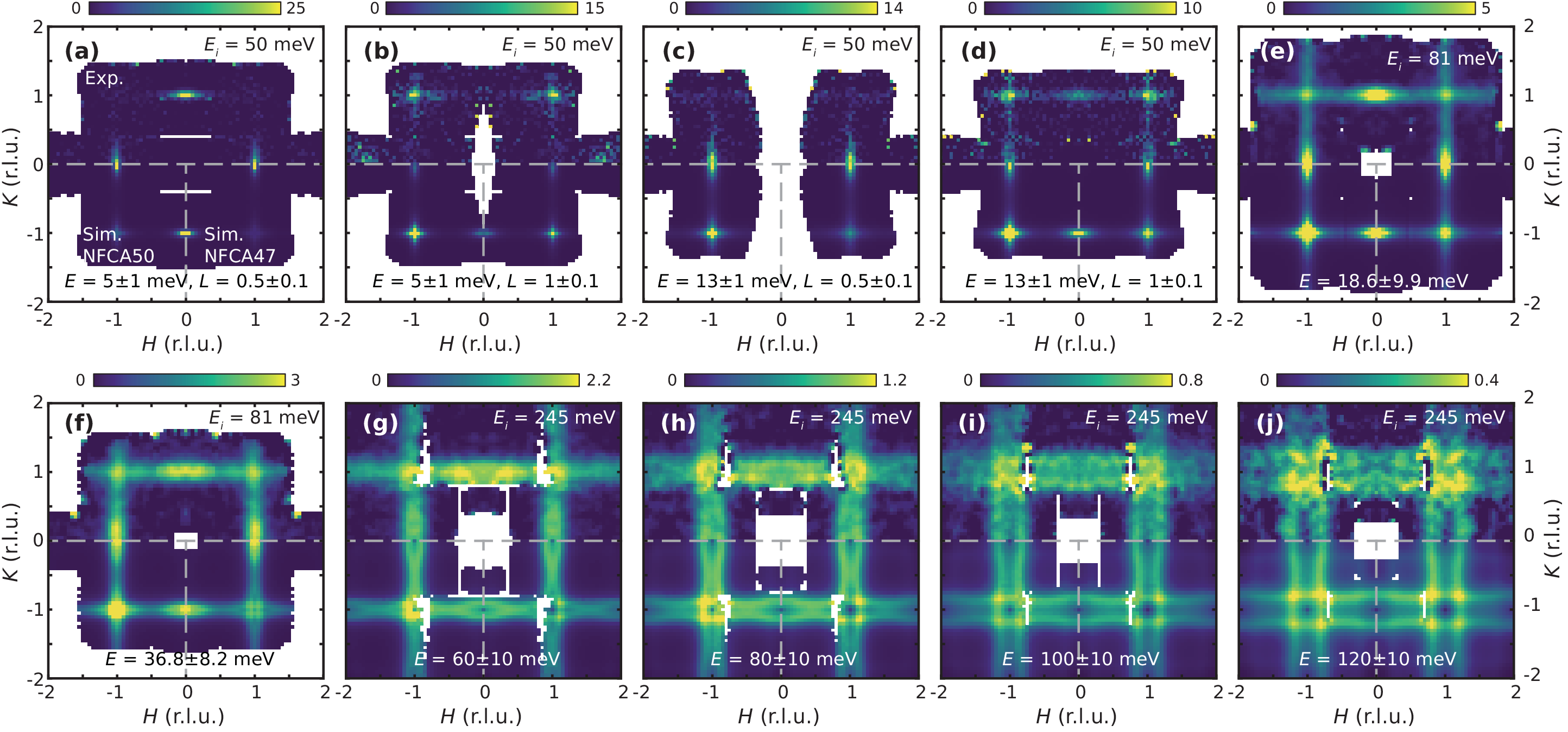}
	\vspace{-12pt} \caption{Constant-energy $HK$-maps for NFCA47 at $T=5$~K, shown in the upper half of each panel. (a)-(d) are \ys{from $E_{\rm i}=50$~meV sample rotations scans}. (e)-(j) \ys{are from $k_{\rm i}\parallel c$ data}. The intensities are color-encoded, and are in units of mbarn~meV$^{-1}$~sr$^{-1}$~f.u.$^{-1}$, where f.u. stands for formula unit. The data are folded into a single quadrant and reproduced in other quadrants. Simulated intensities are shown in the bottom half, for NFCA50 on the left, and for NFCA47 on the right, based on a Heisenberg model for NFCA50 \cite{SI}.
	} 
	
	\vspace{-12pt}
	\label{HKmaps}
\end{figure*}

Single crystals of NFCA47 were synthesized using the self-flux method \cite{Song2016,YSong2021}. 4.3~g of samples were co-aligned, and neutron scattering measurements were carried out using the MAPS chopper spectrometer \cite{Ewings2016} at the ISIS Neutron and Muon source, and the Polarized Triple-Axis Spectrometer (HB-1) at the High-Flux Isotope Reactor (HFIR), Oak Ridge National Laboratory (ORNL). Momentum transfers ${\bf Q}=\frac{2 \pi}{a}H{\bf \hat{a^*}}+\frac{2 \pi}{b}K{\bf \hat{b^*}}+\frac{2 \pi}{c}L{\bf \hat{c^*}}$ are referenced in reciprocal lattice units $(H,K,L)$, where $a\approx b\approx 5.70$~{\AA} and $c\approx6.88$~{\AA} [Fig.~\ref{Schematic}(a)]. The MAPS measurements were carried out \ys{in the $[H0L]$ plane, with} $k_{\rm i}\parallel c$ (incident energy $E_{\rm i}=50$~meV, 81~meV and 245~meV) \ys{or via sample rotation} about the vertical axis ($K$) in steps of 2$^{\circ}$ ($E_{\rm i}=50$~meV). The energy transfer $E$ and $L$ are coupled in the $k_{\rm i}\parallel c$ measurements, whereas four-dimensional $E-{\bf Q}$ data are obtained in the sample rotation measurements, and are analyzed using \texttt{Horace} \cite{Ewings2016}. The measured intensities are converted into absolute units through measurements of a vanadium reference.

The in-plane crystal and magnetic structures of NaFeAs and NFCA50 are compared in Figs.~\ref{Schematic}(b) and (c)\ys{, with the associated scattering represented by red symbols in Figs.~\ref{Schematic}(d) and (e)}. In NaFeAs, the Fe atoms form a square lattice with stripe magnetic ordering at ${\bf Q_{\rm S}}=(1,0)$ \cite{Dai_RMP}. In NFCA50, Fe and Cu atoms order into chains and form an atomic analogue of the stripe magnetic order, leading to superstructure peaks at $(1,0)$ and equivalent positions. The Fe spins in NFCA50 exhibit AF order along the Fe chains, resulting in magnetic peaks at $(0,1)$. Equivalent magnetic peaks occur at integer $(H,1)$ positions, including the {\Neel} vector ${\bf Q}_{\rm N}=(1,1)$. \ys{For the $(0,1)$ magnetic peak, the longitudinal direction $K$ is along the Fe chains, and the transverse direction $H$ is perpendicular to the chains.}
As single crystals of both NaFeAs and NFCA50 are typically twinned, experimentally measured scattering patterns exhibit fourfold rotational symmetry, such that magnetic peaks are seen at ${\bf Q}_{\rm S}=(1,0)/(0,1)$ in both two compounds, and also at ${\bf Q}_{\rm N}=(1,1)$ in NFCA50 [Figs.~\ref{Schematic}(d) and (e)]. See Supplemental Materials for a detailed comparison of expected Bragg peak positions in NaFeAs and NFCA50 \cite{SI}.   

By considering the interlayer atomic and magnetic orderings in NFCA50 [Fig.~\ref{Schematic}(a)], magnetic peaks are expected at half-integer-$L$ ${\bf Q}_{\rm S}$, and at integer-$L$ ${\bf Q}_{\rm N}$. These magnetic peaks are confirmed in the MAPS data of NFCA47 at 5~K [Figs.~\ref{Schematic}(f) and (g)]. The temperature dependence of magnetic peaks in NFCA with different Cu concentrations are compared in Fig.~\ref{Schematic}(h), showing that magnetic order onsets below $T_{\rm N}\sim210$~K in NFCA47, slightly higher than the $x=0.39$ and $0.44$ samples, consistent with an ideal ordered  phase at $x=0.5$.

$HK$-maps of spin excitations in NFCA47 at $T=5$~K are shown in the upper half of panels in Fig.~\ref{HKmaps}. 
For $E=5$~meV, low-energy spin excitations are observed at half-integer-$L$ ${\bf Q}_{\rm S}$ [Fig.~\ref{HKmaps}(a)] and integer-$L$ ${\bf Q}_{\rm N}$ [Fig.~\ref{HKmaps}(b)], consistent with spin waves stemming from the magnetic Bragg peaks [Figs.~\ref{Schematic}(f)-(g)]. Different from the magnetic Bragg peaks that are nearly isotropic in the $HK$-plane, the low-energy spin excitations at ${\bf Q}_{\rm S}$ exhibit a pronounced \ys{elongation perpendicular (transverse) to ${\bf Q}_{\rm S}$} [Figs.~\ref{HKmaps}(a)-(d)].

\begin{figure}
\includegraphics[angle=0,width=0.49\textwidth]{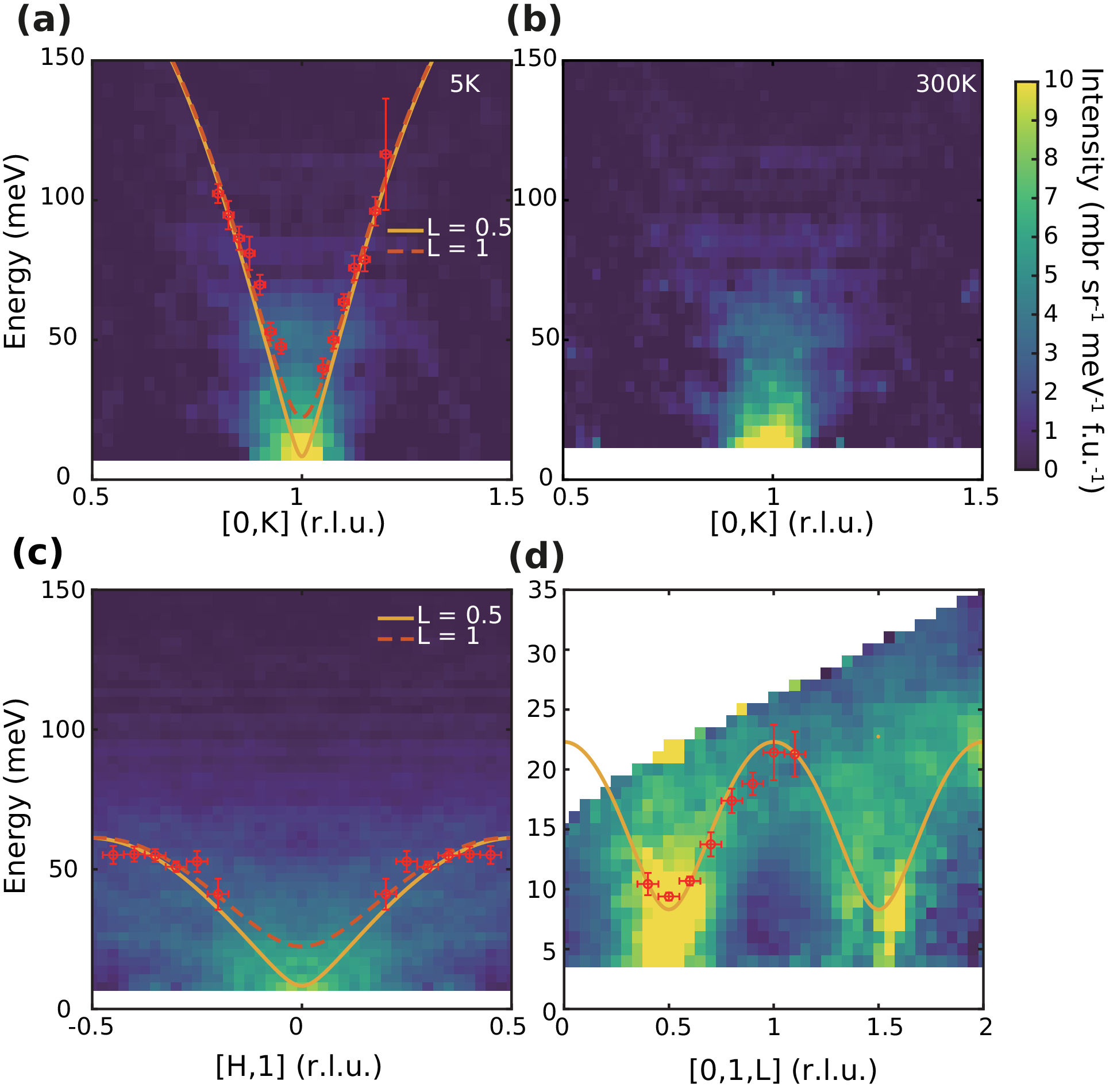}
\vspace{-12pt} \caption{\label{Figure3} $E-\bf{Q}$ slices for NFCA47 with $\bf{Q}$ along $[0,K]$ at (a) 5~K and (b) 300~K, (c) along $[H,1]$ at 5~K, and (d) along $[0,1,L]$ at 5~K. \ys{(a)-(c) are from $E_{\rm i}$=245~meV $k_{\rm i}\parallel c$ measurements}, by binning $H=[-0.2,0.2]$ in (a) and (b), and $K=[0.9,1.1]$ in (c). Data in (d) are from \ys{$E_{\rm i}=50$~meV sample rotation scans}, by binning $H=[-0.1,0.1]$ and $K=[0.9,1.1]$. The red circles are extracted from fits to constant-$\bf{Q}$ cuts, with the horizontal and vertical bars corresponding to the integration range and fit uncertainties. The curves in (a), (c) and (d) are the spin wave dispersion for a Heisenberg model \cite{SI}.
A similar figure showing intensity$\times$energy is reproduced in the Supplemental Material \cite{SI}.}
\label{Fig3}
\vspace{-12pt}
\end{figure}

With increasing energy transfer, the excitations at ${\bf Q}_{\rm S}$ further broaden along the transverse direction, become connected to excitations at ${\bf Q}_{\rm N}$, and eventually becoming independent of the transverse direction for $E\gtrsim60$~meV [Figs.~\ref{HKmaps}(e)-(h)]. For the \ys{in-plane direction along ${\bf Q}_{\rm S}$ (longitudinal direction)}, the excitations broaden with increasing energy at a much slower rate, and a splitting is observed only for $E\gtrsim80$~meV [Figs.~\ref{HKmaps}(h)-(j)]. Such a splitting is also clearly resolved in \ys{Fig.~\ref{Fig3}(a)}, with similar spin excitations persisting in the paramagnetic state [Fig.~\ref{Fig3}(b)] \cite{SI}, consistent with the energy scale of spin correlations well exceeding $T_{\rm N}$. \ys{Similar} persistence of spin excitations well above $T_{\rm N}$ are observed across various FePn/Ch compounds \cite{Dai_RMP}. 

\ys{The strong in-plane anisotropy of excitations is also clearly revealed by comparing $E$-${\bf Q}$ slices with ${\bf Q}$ corresponding to in-plane longitudinal and transverse directions [Figs.~\ref{Fig3}(a) and (c)]. In combination with the $E$-$L$ slice that shows a band top of $\approx20$~meV along $L$ [Fig.~\ref{Fig3}(d)], these $E$-${\bf Q}$ slices demonstrate that spin excitations above $\sim60$~meV are sheets perpendicular to the in-plane longitudinal direction (Fe chain direction), and are thus quasi-1D. Due to orthogonal twin domains, these quasi-1D excitations give rise to grid-like $HK$-maps [Figs.~\ref{HKmaps}(g)-(j)]. Similar grid-like $HK$-maps were found in the heavy fermion metal CeSb$_2$, also due to its quasi-1D spin dynamics and orthogonal twin domains \cite{Shan2025}. The quasi-1D nature of magnetism in NFCA47 suggests the 2D electronic structure close to the Fermi level in ARPES measurements \cite{CEMatt2016} are dominated by As atoms that form a 2D network \cite{SZhang2017}. 
} 


The local susceptibility $\chi^{\prime\prime}(\omega)$ of NFCA47 can be obtained by averaging the measured scattering intensities over momentum \cite{Dai_RMP}, as shown in Fig.~\ref{Schematic}(i). The total fluctuating moment $\langle m^2 \rangle$, related to the integral of $\chi^{\prime\prime}(\omega)$ over energy, is found to be 6.3~$\pm~1.0~\mu_{\rm B}^2$/Fe at 5~K \cite{SI}.
By normalization to a weak structural Bragg peak in HB1 measurements, an ordered moment $M=1.2~\pm~0.1$~$\mu_{\rm B}$/Fe at 5~K is found for NFCA47 \cite{SI}, similar to values in NFCA with $x=0.44$ \cite{Song2016}. The total moment $M_0^2=M^2+\langle m^2\rangle$ is then found to be 7.7~$\pm~1.1~\mu_{\rm B}^2$/Fe. Since $M_0^2=g^2S(S+1)$, the experimental value of $M_0^2$ suggests an effective spin $S\approx1$, significantly larger than $S\approx1/2$ for NaFeAs \cite{CZhang2014}, while similar to $S\approx1$ for FeSe \cite{Wang2016}. 
Analyses of $\chi^{\prime\prime}(\omega)$ for 190~K and 300~K yields $M_0^2~=~6.4~\pm~0.9$ and $6.3~\pm~1.0~\mu_{\rm B}^2$/Fe \cite{SI}, smaller than the value at 5~K, \ys{possibly due to the broadening of spectral weight to higher energies.} 

By fitting constant-${\bf Q}$ cuts at various momenta, the dispersion of spin excitations in NFCA47 is extracted, which can be described by a Heisenberg model containing three exchange couplings [Fig.~\ref{Schematic}(a)], with $SJ_{\rm \parallel}=90.1(3)$~meV, $SJ_{\rm \perp}=-2.4(1)$~meV, and $SJ_c=0.15(5)$~meV [solid lines in Fig.~\ref{Fig3}] \cite{SI}. These exchange couplings are in reasonable agreement with RIXS measurements ($SJ_{\rm \parallel}=79$~meV) \cite{YSong2021} and DFT+$U$ calculations ($SJ_{\rm \parallel}\approx75$~meV, $SJ_{\rm \perp}\approx-1.1$~meV, $SJ_{c}\approx0.35$~meV). The scattering signals at magnetic zone centers peak around $\approx9$~meV, resulting from a single-ion anisotropy term $S\Delta=0.096(5)$~meV \cite{SI}. Interestingly, $SJ_{\rm \parallel}$ in NFCA47 is roughly twice of $SJ_{1a}$ (connecting nearest-neighbor AF spins) in NaFeAs \cite{CZhang2014}, mirroring the difference in $S$. 

The experimental {\Neel} excitations in NFCA47 (top of panels in Fig.~\ref{HKmaps}) are weaker than expected for NFCA50 (bottom left of panels in Fig.~\ref{HKmaps}), which indicates they are vulnerable to Fe impurities on the Cu site, whereas the stripe excitations are less susceptible. The spin excitations in NFCA47 can be quantitatively reproduced using exchange couplings described above, a $\mathbf{Q}$-dependent damping, and an empirical function that suppresses the {\Neel} excitations \cite{SI}, with the simulated intensities shown in the bottom right of panels in Fig.~\ref{HKmaps}. The agreement between experimental and simulated scattering patterns in Fig.~\ref{HKmaps} shows that while biquadratic couplings are likely present in NFCA47, the resulting spin excitations can be mapped to an effective Heisenberg model with the same symmetry as its crystal structure. In contrast, biquadratic couplings are necessary to describe the spin excitations in quasi-2D FePn/Ch, when using Hamiltonians that retain \ys{symmetries of the square lattice} \cite{Wysocki2011,Yu2012,RLiu2024}.

While NFCA47 is an insulator, its spin waves are significantly broadened in energy [Fig.~\ref{Fig3} and Supplemental Fig.~7], different from typical insulating magnets but similar to metallic/superconducting FePn/Ch \cite{Dai_RMP}. The exchange couplings obtained from the dispersion of spin waves and an effective $S\approx1$ suggest that magnetic frustration \cite{Balents2010,Broholm2020} and spin-1/2 spinons \cite{Tennant1993,Lake2013,Schlappa2018,Scheie2022} play no role. Instead, we attribute the broadening to (1) an intermediate intra-orbital Coulomb interaction that lead to both broad excitations and the absence or weakening of a Haldane gap \cite{Jadewska2023}; (2) a charge gap ($\sim20$~meV \cite{CEMatt2016,SZhang2017}) much smaller than the spin wave bandwidth, with charge excitations across such a small a gap leading to damping effects absent in typical insulating magnets; (3) disorder effects due to deviations from the ideal NFCA50 structure, with end-chain defects leading to coexistent cluster spin-glass and long-range AF order \cite{YXin2020}. 



There are several remarkable similarities in the spin dynamics of NFCA47 and FeSe \cite{Wang2016,RLiu2024}: (1) magnetic excitations stem from both the stripe ${\bf Q}_{\rm S}$ and {\Neel} ${\bf Q}_{\rm N}$ vectors, (2) excitations from $\bf{Q}_{\rm S}$ exhibit a pronounced transverse elongation, (3) {\Neel} excitations are weaker relative to stripe excitations, and (4) the effective spin is $S\approx1$. 
It is worth noting that while (1) is also found in Mn-doped BaFe$_2$As$_2$, the ${\bf Q}_{\rm S}$ and ${\bf Q}_{\rm N}$ excitations there have distinct origins \cite{GTucker2012}, different from NFCA47. 

\begin{figure}
	\includegraphics[angle=0,width=0.49\textwidth]{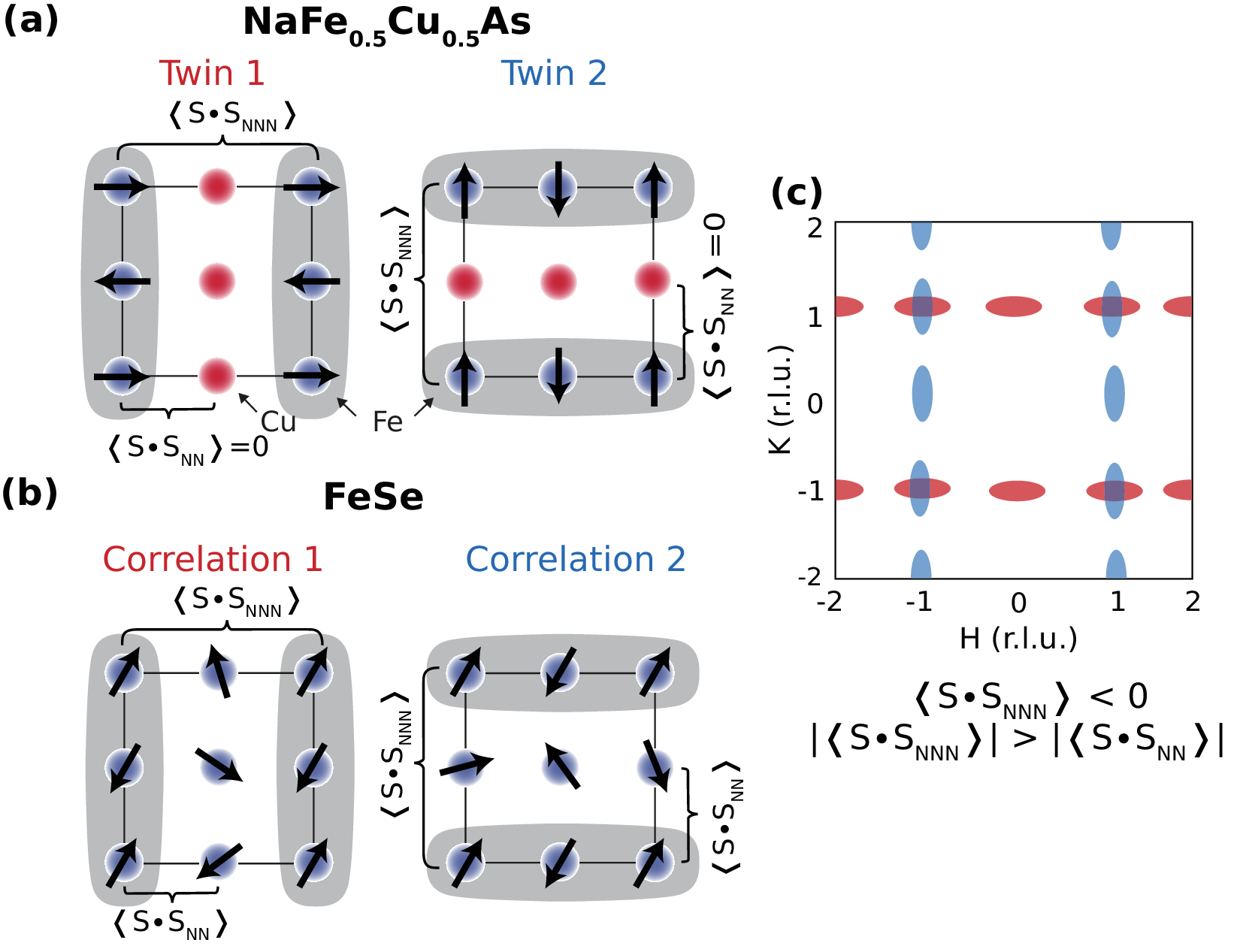}
	\vspace{-12pt} \caption{(a) Schematic configurations for the two twins of NFCA50. (b) Schematics for two kinds of spin correlations proposed for FeSe. In both NFCA50 and FeSe, the spin correlations are dominantly quasi-1D, represented by the shaded stripe-like areas. The NNN interchain correlations are stronger than the NN interchain correlations in magnitude ($|\langle S \cdot S_{\rm NNN}\rangle|>|\langle S \cdot S_{\rm NN}\rangle|$). (c) Schematic representation of excitations from (a) or (b), with the two colors corresponding to scattering patterns arising from the two structural twins in (a) or the two kinds of dynamic spin correlations in (b).}
	\vspace{-12pt}
	\label{FeSeCompare}
\end{figure}

As the spin excitations in NFCA47 are reasonably described by a model of weakly coupled antiferromagnetic chains [Fig.~\ref{Fig3}], the similarities described above suggest the dynamic spin correlations in FeSe may likewise have a quasi-1D character. First-principles calculations reveal a number of proximate ground states in FeSe that order along ${\bf Q}_{\rm S}$-${\bf Q}_{\rm N}$, including stripe and staggered dimer, trimer and tetramer orders, accounting for the unusual paramagnetic ground state of FeSe \cite{Glasbrenner2015}. These orders all consist of AF chains with distinct interchain configurations, and their near-degeneracy suggest dominant 1D spin correlations which experimentally manifest as a prominent transverse elongation of low-energy magnetic excitations \cite{Wang2016}. 
In theoretical models that account for the paramagnetic state of FeSe via magnetic frustration, the spin dynamics are also expected to exhibit quasi-1D features \cite{FWang2015,She2018}. 

In NFCA50 (approximated by our measurements), as the Cu chains are nonmagnetic, the nearest-neighbor (NN) interchain spin correlations between the Fe chains and Cu chains are zero. The dominant interchain correlations at low energies are then the ferromagnetic next-nearest-neighbor (NNN) correlations between Fe chains, which in  combination with the twinning of two structural domains, lead to low-energy excitations at ${\bf Q}_{\rm S}$ and ${\bf Q}_{\rm N}$ [Figs.~\ref{FeSeCompare}(a) and (c)]. Given the similarities of spin excitations in NFCA50 and FeSe, a similar explanation for low-energy excitations in FeSe can be proposed \cite{SI}: although the Fe atoms in FeSe form a square lattice, the dynamic spin correlations are quasi-1D with weak NN interchain correlations but relatively strong ferromagnetic NNN interchain correlations, and two variants of such excitations related by a 90$^{\circ}$ rotation give rise to scattering similar to that in NFCA50 [Figs.~\ref{FeSeCompare}(b) and (c)]. It should be noted that while one variant of quasi-1D excitations is anticipated in single-domain NFCA50, the two variants coexist in detwinned orthorhombic FeSe \cite{Chen2019,RLiu2024}, resulting from the much weaker anisotropy of its nematic state, compared to the Fe-Cu chains in NFCA50. Compared to the ideal NFCA50, the weakened {\Neel} excitations in NFCA47 suggests that the ratio between {\Neel} and stripe excitations in NFCA can be tuned by Fe/Cu concentration, reminisicent of their competition in FeSe upon changing temperature \cite{Wang2016}. 


It is interesting to note that in contrast to NFCA, heavily Cu-substituted Ba(Fe$_{1-x}$Cu$_x$)$_2$As$_2$ (BFCA) only exhibits short-range ${\bf Q}_{\rm S}$ magnetic order, with neither significant atomic ordering nor ${\bf Q}_{\rm N}$ magnetic order \cite{WWang2017}. At the same time, NaFeAs exhibits ${\bf Q}_{\rm N}$ excitations for $E\gtrsim40$~meV \cite{CZhang2014,RLiu2024}, a feature absent in most FePn/Ch compounds such as BaFe$_2$As$_2$ \cite{LHarriger2011}. These findings indicate that the presence(absence) of low-lying ${\bf Q}_{\rm N}$ excitations in NaFeAs(BaFe$_2$As$_2$) is correlated with the presence(absence) of ${\bf Q}_{\rm N}$ magnetic order and atomic ordering in heavily Cu-substituted NFCA(BFCA), which implicates a contribution of magnetic interactions in the formation of atomic orderings in magnetically diluted FePn/Ch compounds. The theoretical capability to predict atomic orderings realized upon magnetic dilution in FePn/Ch (and other quantum materials more broadly) will be valuable, and considering the sizable energy scale of magnetic excitations ($\sim200$~meV) in these materials, the inclusion of magnetic contributions to the formation energies may be required.

Our findings conclusively show that despite the planar crystal morphology and 2D electronic states just below the Fermi level \cite{CEMatt2016}, the magnetic excitations in NFCA47 are unequivocally quasi-1D. This in turn suggests magnetic dilution as a promising approach for realizing quasi-1D FePn/Ch systems. Such an effort proved to be fruitful in (Fe$_{1-x}$Cu$_x$)$_{1+y}$Te, where Fe-Cu ordered phases with Fe chains or ladders were suggested \cite{SCao2024}.  The layered structure of these quasi-1D FePn/Ch compounds enables a broad range of spectroscopic studies \cite{CYe2015,CEMatt2016,ACharnukha2017,Zhang2018,YXin2020,YSong2021}, paving the way for synergistic combinations of numerically exact many-body theories and spectroscopy measurements in quasi-1D FePn/Ch, as carried out for the cuprates \cite{Dagotto1996,Walters2009,chain_Cuprate_Science}. 




This work is supported by the National Key R\&D Program of China (Grant No. 2024YFA1409200), the National Natural Science Foundation of China (Grants No. 12274363, No. 12334008, No. 52271037), the Fundamental Research Funds for the Central Universities (Grant No. 226-2024-00068), and the Key Research Program of Chinese Academy of Sciences (No. ZDBS-ZRKJZ-TLC021).The neutron scattering and materials synthesis work at Rice university are supported by the U.S. Department of Energy (DOE), Basic Energy Sciences (BES), under Contracts No. DE-SC0012311, No. DE-SC0026179, and the Robert A. Welch Foundation, Grant No. C-1839 (P.D.), respectively. Experiments at the ISIS Neutron and Muon Source were supported by beam time allocations RB1610394 and RB1810055 from the Science and Technology Facilities Council. Data is available at: \url{https://doi.org/10.5286/ISIS.E.RB1610394} and \url{https://doi.org/10.5286/ISIS.E.RB1810055}. A portion of this research used resources at the High Flux Isotope Reactor, DOE Office of Science User Facilities operated by the Oak Ridge National Laboratory. The beam time was allocated to HB-1 on proposal number IPTS-19995.1.

\bibliography{NFCA_INS.bib}

\end{document}